\documentclass[reprint, showpacs]{revtex4-1}
\usepackage{graphicx}
\usepackage{amsfonts}
\usepackage{natbib}
\usepackage{float}

\newcommand{\be}{\begin{equation}}
\newcommand{\ee}{\end{equation}}
\newcommand{\bea}{\begin{eqnarray}}
\newcommand{\eea}{\end{eqnarray}}

\newcommand{\ep}{\varepsilon}
\newcommand{\al}{\alpha}

\begin{document}
\title{The level repulsion exponent of localized chaotic  eigenstates as a function of the classical transport
  time scales in the stadium billiard}

\author{Benjamin Batisti\'c}
\author{\v Crt Lozej}
\author{Marko Robnik}

\affiliation{CAMTP - Center for Applied Mathematics and Theoretical
Physics, University of Maribor, Mladinska 3, SI-2000 Maribor, Slovenia, European Union}

\date{\today}

\begin{abstract}
  We study the aspects of quantum localization in the stadium billiard, which is a classically
  chaotic ergodic system, but in the regime of slightly distorted circle billiard
  the diffusion in the momentum space is very slow. In quantum systems with discrete energy spectrum
  the Heisenberg time $t_H =2\pi \hbar/\Delta E$, where $\Delta E$ is the mean level spacing (inverse
  energy level density), is an important time scale. The classical transport time scale $t_T$
  (diffusion time) in relation to the Heisenberg time scale $t_H$ (their ratio is the parameter
  $\alpha=t_H/t_T$) determines the degree of localization of the chaotic eigenstates, whose measure $A$
  is based on the information entropy. The localization of chaotic eigenstates is reflected also in the
  fractional power-law repulsion between the nearest energy levels in the
  sense that the probability density (level spacing distribution)
  to find successive levels on a distance $S$ goes like $\propto S^\beta$ for
  small $S$, where $0\leq\beta\leq1$, and $\beta=1$ corresponds to completely
  extended states. We show that the level repulsion exponent $\beta$ is a unique rational
  function of $\al$, and $A$ is a unique rational function of $\al$.
  $\beta$  goes from $0$ to $1$ when $\al$ goes from $0$ to $\infty$. Also, $\beta$
  is a linear function of $A$, which is similar as in the quantum kicked rotator, 
  but different from a mixed type billiard.
\end{abstract}


\maketitle

\section{Introduction}
\label{sec1}

In quantum chaos we study phenomena in the quantum domain, or in
other wave systems described also by the wave equations different from the
Schr\"odinger equation, which correspond to the classical chaos in the
semiclassical limit (short wavelength approximation) \cite{Stoe,Haake}.
The quantum localization of classical chaotic
diffusion in the time-dependent domain is one of the most important fundamental
phenomena in quantum chaos, discovered and studied first in the quantum 
kicked rotator \cite{Cas1979,Chi1981,Chi1988,Izr1990} by Chirikov, Casati,
Izrailev, Shepelyansky, Guarneri and many others. An excellent
extensive account and review has been given by Izrailev
\cite{Izr1988,Izr1989,Izr1990}. This type of phenomena is generic in
chaotic Floquet (time-periodic) Hamilton systems.

In the time-independent domain the quantum localization 
is manifested in the localized chaotic eigenstates. In the case of
the quantum kicked rotator, for example, one sees the exponentially
localized eigenstates in the dimensionless space of the angular momentum
quantum number. For an extensive review see \cite{Izr1990}.
This phenomenon is closely related to the Anderson
localization in one dimensional disordered lattices as shown
for the first time by Fishman, Grempel and Prange \cite{FGP1982},
and later discussed and studied by many others \cite{Stoe,Haake}.

The quantum localization in billiards has been reviewed by
Prosen in reference  \cite{Pro2000}. Here we have to look
at the localization properties of the localized chaotic eigenstates
in the quantum phase space, which means study of the Wigner functions
(which are real valued but not positive definite), or better,
the Husimi functions, which are real and positive definite, and can
be treated as a quasi-probability density. In the semiclassical limit
we are interested in the quantum-classical correspondence of these
structures in the phase spaces.

Recently, we \cite{BatRob2013A} have studied the localization of chaotic
eigenstates in the mixed-type billiard \cite{Rob1983,Rob1984},
after the separation of the
chaotic and regular eigenstates based on such quantum-classical
correspondence \cite{BatRob2013B}. We have introduced two localization
measures, one based on the information entropy denoted by $A$ and
used in this paper, and the other one $C$ based on the correlations.
We have shown that $A$ and $C$ are linearly related and thus equivalent.

In this paper we study localization
properties of eigenstates in the stadium billiard of Bunimovich \cite{Bun1979},
which is a chaotic ergodic system.
Studies of the slow diffusive regime in this system and the related
quantum localization were initiated in Ref. \cite{BCL1996},
while the detailed aspects of classical diffusion have been
investigated in our recent paper \cite{LozRob2018A}.

Another fundamental phenomenon in quantum chaos in the time-independent
domain is the statistics of the fluctuations in the energy spectra. In analogy
with the time-periodic systems we find functional relationship between
the localization measure $A$ and the spectral (energy) level
repulsion exponent $\beta$, to be precisely defined below. 
$A$ and $\beta$ are unique functions of the parameter $\al=t_H/t_T$, which by
definition is the ratio of two most important time scales in the
system, namely the Heisenberg time $t_H=2\pi \hbar/\Delta E$, where $\Delta E$
is the mean energy level spacing, and the classical transport (diffusion)
time scale $t_T$. These findings are the main result of this work.

The statistical properties of energy spectra of quantum systems are
universal \cite{Stoe,Haake,Mehta,GMW,Rob1998}.
In the sufficiently deep semiclassical limit (when $\al$ is large enough,
$\al \gg 1$, which can always been achieved by sufficiently small
effective $\hbar$) and in general mixed type systems,
they are determined solely by the type of classical motion,
which can be either regular or chaotic \cite{Percival1973,BerRob1984,Rob1998,BatRob2010,
BatRob2013A,BatRob2013B}.
The level statistics is Poissonian if the underlying classical invariant
component is regular, whilst for chaotic extended states
the Random Matrix Theory (RMT) applies \cite{Mehta}, 
specifically the Gaussian Orthogonal Ensemble statistics
(GOE) in case of an antiunitary symmetry. This is the {\em Bohigas-Giannoni-Schmit
conjecture} \cite{Cas1980, BGS1984}, which has been proven only recently 
\cite{Sieber,Mueller1,Mueller2,Mueller3,Mueller4} using the semiclassical methods
and the periodic orbit theory developed around 1970 by Gutzwiller
(\cite{Gutzwiller1980} and the references therein), an approach initiated by 
Berry \cite{Berry1985}, well reviewed in \cite{Stoe,Haake}.

The classification regular-chaotic can be done by analyzing the
structure of eigenstates in the quantum phase space, based on the
Wigner functions, or Husimi functions \cite{BatRob2013B}. Of course, in the stadium
billiard all eigenstates are of the chaotic type, but can be strongly localized
if $\al$ is small enough, $\al\ll 1$.

The most important statistical measure
is the level spacing distribution $P(S)$, assuming spectral
unfolding such that $\left<S\right>=1$. For integrable 
systems  and regular levels of mixed type systems
$P(S)=\exp\left(-S\right)$, whilst for extended chaotic systems it
is well approximated by the Wigner distribution 
$P(S)= \frac{\pi S}{2}\exp\left(-\frac{\pi}{4}\,S^2\right)$.
The distributions differ significantly in a small $S$ regime, where there
is no level repulsion in a regular system and a linear level repulsion,
$P(S)\propto S$, in a chaotic system. Localized chaotic states exhibit 
the fractional power-law level repulsion $P(S)\propto S^\beta$, as clearly
demonstrated recently by Batisti\'c and Robnik \cite{BatRob2010,BatRob2013A,BatRob2013B}.

The localization is a pure quantum effect which appears if the Heisenberg
time $t_H$,  which is the time scale on which the 
quantum evolution follows the classical one, is smaller than the relevant 
classical transport time $t_T$ (diffusion or ergodic time).
Up to the Heisenberg time the quantum system behaves as if the
evolution operator had a continuous spectrum, but at times longer
than Heisenberg time the discrete spectrum of the evolution
operator becomes resolved, and the interference effects set in, 
resulting in a destructive interference causing the quantum localization.
Thus the parameter $\al=t_H/t_T$ plays a key role.
The ergodic time may be very long, especially if the chaotic region has
a complicated, but typical KAM structure, due to the presence of the
partial barriers in the form of barely destroyed irrational tori, called cantori, 
which allow for a very slow transport only. However, in this paper
we study the stadium, which is ergodic, so no KAM structures are present,
although cantori can be and in fact are present.

The weak ($\beta<1$) level repulsion of localized states is empirically
observed, but the whole distribution $P(S)$ is globally theoretically not known.
Several different distributions which would extrapolate the small
$S$ behaviour were proposed. The most popular are the Izrailev
distribution \cite{Izr1988,Izr1989,Izr1990} and the Brody distribution
\cite{Bro1973,Bro1981}. 
The Brody distribution is a simple generalization of the Wigner distribution. 
Explicitly, the Brody distribution is

\be \label{BrodyP}
P_B(S) = c S^{\beta} \exp \left( - d S^{\beta +1} \right), \;\;\; 
\ee
where 

\be \label{Brodyab}
c = (\beta +1 ) d, \;\;\; d  = \left( \Gamma \left( \frac{\beta +2}{\beta +1}
 \right) \right)^{\beta +1}
\ee
with  $\Gamma (x)$ being the Gamma function. It interpolates the
exponential and Wigner distribution as $\beta$ goes from $0$ to $1$. 
The Izrailev distribution is a bit more complicated but has the feature of being
a better approximation for the GOE distribution at $\beta=1$. One 
important theoretical plausibility argument by Izrailev in support of
such intermediate level spacing distributions is that
the joint level distribution of Dyson circular ensembles can be extended
to noninteger values of the exponent $\beta$ \cite{Izr1990}.
However, recent numerical results show that Brody distribution 
is slightly better in describing real data
\cite{BatRob2010,BatRob2013A,ManRob2013,BatManRob2013}, 
and is simpler, which is the reason why we prefer and use it.

The open question is how does the level repulsion parameter
$\beta$ depend on the localization. 
This question was raised for the first time by Izrailev
\cite{Izr1988,Izr1989,Izr1990},
where he numerically studied the quantum kicked rotator, which is a 
1D time-periodic system. His result showed that the parameter $\beta$, which
was obtained using the Izrailev distribution, is functionally related to the
localization measure defined as the information entropy of the
eigenstates in the angular momentum representation. His results were
recently confirmed and extended, with the much greater numerical accuracy
and statistical significance \cite{ManRob2013,BatManRob2013}.
Moreover, in Ref. \cite{BatRob2013A} it has been demonstrated that
$\beta$ is a unique function of $A$ in the billiard with the mixed phase space
\cite{Rob1983,Rob1984}. 

In this paper we show that there is indeed a functional
relation between the level repulsion parameter $\beta$ and the localization 
measure $A$ also in the stadium billiard, in analogy with the
quantum kicked rotator and the above mentioned billiard, but the
functional form is different. We also show that $\beta$ is a unique function
of $\al$.

\section{The billiard systems and Poincar\'e-Husimi functions}
\label{sec2}

\begin{figure}[H]
  \centering
  \includegraphics{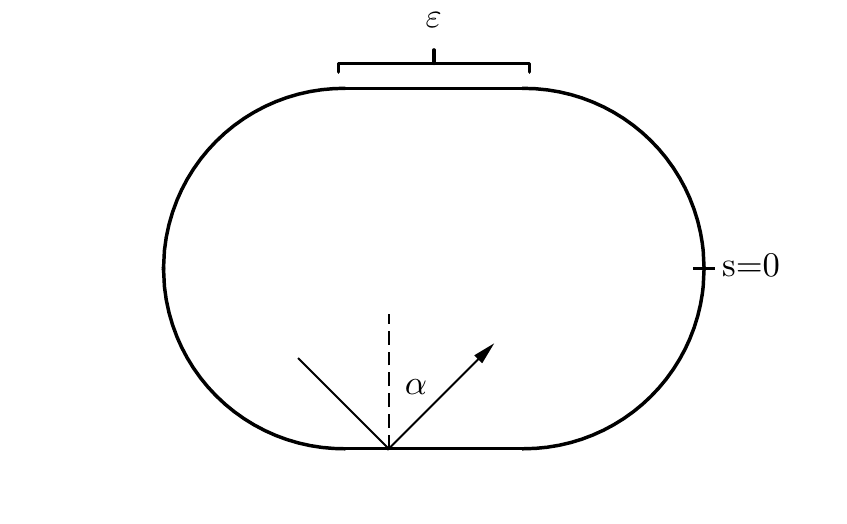}
  \caption{The geometry and notation of the stadium billiard of Bunimovich.}
  \label{figlr1}
\end{figure}

The stadium billiard \cite{Bun1979} is defined as two semicircles of
radius 1 connected by two parallel straight lines of lentgh $\ep$, as shown
in Fig. \ref{figlr1}.

For a 2D billiard the most natural coordinates in the phase space
$(s,p)$ are the arclength $s$ round the billiard boundary, $s\in [0,{\cal L}]$,
where ${\cal L}$ is the circumference, and the sine of
the reflection angle, which is the component of the unit velocity
vector tangent to the boundary at the collision point, equal to $p=\sin \al$,
which is the canonically conjugate momentum to $s$. These are the
Poincar\'e-Birkhoff coordinates. The bounce map $(s_1,p_1)
\rightarrow (s_2,p_2)$ is area preserving, and the phase portrait
does not depend on the speed (or energy) of the particle. Quantum mechanically
we have to solve the stationary Schr\"odinger equation, which in a
billiard is just the Helmholtz equation $\Delta \psi + k^2 \psi =0$
with the Dirichlet boundary conditions  $\psi|_{\partial {\cal B}}=0$.
The energy is $E=k^2$. The important quantity is
the boundary function 

\be  \label{BF}
u(s) = {\bf n}\cdot \nabla_{{\bf r}} \psi \left({\bf r}(s)\right),
\ee
which is the normal derivative of the wavefunction $\psi$ at the 
point $s$ (${\bf n}$ is the unit normal vector). 
It satisfies the integral equation

\be \label{IEBF}
u(s) = -2 \oint dt\; u(t)\; {\bf n}\cdot\nabla_{{\bf r}} G({\bf r},{\bf r}(t)),
\ee
where $G({\bf r},{\bf r'}) = -\frac{i}{4} H_0^{(1)}(k|{\bf r}-{\bf r'}|)$ is
the Green function in terms of the Hankel function $H_0(x)$. It is important
to realize that boundary function $u(s)$ contains complete information
about the wavefunction at any point ${\bf r}$ inside the billiard by the equation

\be \label{utopsi}
\psi_m({\bf r})  = - \oint dt\; u_m(t)\; G\left({\bf r},{\bf r}(t)\right).
\ee
Here $m$ is just the index (sequential quantum number) of the $m$-th eigenstate.
Now we go over to the quantum phase space. We can calculate the Wigner
functions \cite{Wig1932} based on $\psi_m({\bf r})$. However, in billiards it is advantageous to
calculate the Poincar\'e - Husimi functions. The Husimi functions \cite{Hus1940} are
generally just Gaussian smoothed Wigner functions. Such smoothing makes
them positive definite, so that we can treat them somehow as quasi-probability 
densities in the quantum phase space, and at the same time we eliminate the
small oscillations of the Wigner functions around the zero level, which do
not carry any significant physical contents, but just obscure the picture.
Thus, following  Tualle and Voros \cite{TV1995} and B\"acker et al
\cite{Baecker2004}, we introduce \cite{BatRob2013A,BatRob2013B}. 
the properly ${\cal L}$-periodized coherent states
centered at $(q,p)$, as follows

\bea \label{coherent}
c_{(q,p),k} (s) & =  & \sum_{m\in {\bf Z}} 
\exp \{ i\,k\,p\,(s-q+m{\cal L})\}  \times \\ \nonumber
 & \exp & \left(-\frac{k}{2}(s-q+m{\cal L})^2\right). 
\eea
The Poincar\'e - Husimi function is then defined as the absolute square
of the projection of the boundary function $u(s)$ onto the coherent
state, namely

\be \label{Husfun}
H_m(q,p) = \left| \int_{\partial {\cal B}} c_{(q,p),k_m} (s)\;
u_m(s)\; ds \right|^2.
\ee
The {\em entropy localization measure} denoted by $A$ is defined as

\be \label{locA}
A = \frac{\exp \left<I\right>}{N_c},
\ee
where

\be  \label{entropy}
I = - \int dq\, dp \,H(q,p) \ln \left((2\pi\hbar)^f H(q,p)\right)
\ee
is the information entropy.  Here $f$ is the number of degrees
of freedom (for 2D billiards $f=2$) and $N_c$ is a number of cells on the 
classical chaotic domain, $N_c=\Omega_c/(2\pi\hbar)^f$, where
$\Omega_c$ is the classical phase space volume of the classical chaotic component.
The mean $\left<I\right>$ is obtained by
averaging $I$ over a sufficiently large number of consecutive chaotic
eigenstates.
In the case of uniform distribution (extended eigenstates) $H=1/\Omega_C={\rm const.}$
the localization measure is $A=1$, while in the case of the strongest localization
$I=0$, and $A=1/N_C \approx 0$.
The Poincar\'e - Husimi function $H(q,p)$
(\ref{Husfun}) (normalized) was calculated on the grid points $(i,j)$
in the phase space $(s,p)$.
We express the localization measure in terms of the discretized Husimi
function.
In our numerical calculations we have put $2\pi\hbar=1$, and
thus $H_{ij}=1/N_C$ in case of extendedness, while for maximal localization
$H_{ij}=1$ at just one point, and zero elsewhere.

To get a good estimate of $\beta$ we need many more levels (eigenstates)
than in calculating $A$. 
The parameter $\beta$ was computed for 40 diffrent values of the parameter
$\epsilon$: $\epsilon_j = 0.01 + 0.0025\,j$ where $j \in [0,1..39]$ and on
12 intervals in $k$ space: $(k_i,k_{i+1})$ where $k_i = 500 + 290\,i$ and $i\in[0,1..11]$.
This is $40\times12=480$ values of $\beta$ altogether. More than $4\times10^6$
energy levels were computed for each $\epsilon$. The size of the intervals
in $k$ was chosen to be maximal and such that the Brody distribution gives
a good fit to the level spacing distributions of the levels in the intervals.

For each $\beta(\epsilon_j,(k_i,k_{i+1}))$ an associated localization measure $A$
was computed on a sample of 1000 consecutive levels around $\bar{k}_i = (k_i + k_{i+1})/2$,
which is a mean value of $k$ on the interval $(k_i, k_{i+1})$.

\begin{figure}[H]
  \centering
  \includegraphics[width=9cm]{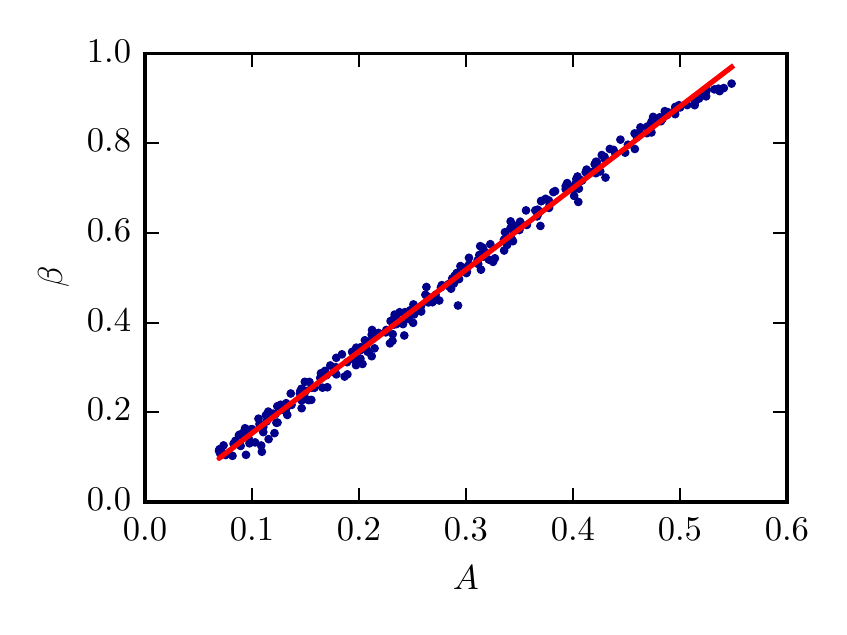}
  \caption{The level repulsion exponent $\beta$ as a function of the
    entropy localization measure $A$ for variety of stadia of
    different shapes $\ep$ and energies $E=k^2$.}
  \label{betaVsA}
\end{figure}

The almost linear dependence of $\beta$ on $A$ is shown 
in Fig. \ref{betaVsA}. 
This is very similar to the case of the quantum kicked rotator
\cite{Izr1990,ManRob2013,BatManRob2013}. In both cases the scattering
of points around the mean linear behaviour is significant, and
probably it is related to the fact that the localization measure
of eigenstates has some distribution, as observed and discussed in
Ref. \cite{ManRob2015}.
There is a great lack in theoretical understanding of the 
physical origin of this phenomenon, even in the case of (the long standing research on) 
the quantum kicked rotator, 
except for the intuitive idea, that energy spectral properties should be 
only a function of the degree of localization, because the localization
gradually decouples the energy eigenstates and levels, switching the linear
level repulsion $\beta=1$ (extendedness) to a power law
level repulsion with  exponent $\beta < 1$ (localization). 
The full physical explanation is open for the future.

\section{The role of the classical transport time scales}
\label{sec3}

As explained in the introduction the role of classical transport time scale $t_T$
is importnat in the semiclassical limit (short wavelength approximation), in relation
to the Heisenberg time scale  $t_H$. We define the parameter $\al=t_H/t_T$,
which controls the quantum localization phenomenon. Usually, as in Ref. \cite{BatRob2013B},
$t_T$ is definesd as the time at which an ensemble of  initial conditions in
the momentum space with initial zero variance of its Dirac delta distribution
reaches a certain fraction of the asymptotic value. In the stadium billiard
for small $\ep$ we have a diffusive regime and thus $t_T$ can be defined
as the diffusion time extracted from the exponential approach of
the momentum variance to the asymptotic value, as has been recently carefully
studied in Ref.  \cite{LozRob2018A}.

For the sake of completeness we derive here the formula for $\al$ in billiards,
following the presentation in Ref. \cite{BatRob2013B}.
In billiards the transport time can be also defined in terms of the number
of collisions (bounces, or iterations of the bounce map), the discrete number
$N_T$.

Let us consider the Heisenberg time and the classical
transport time for a chaotic billiard. According to the
leading order of the Weyl formula, which is in fact just the simple
Thomas-Fermi rule, we have for the number of levels $N(E)$ below
and up to the energy $E$ of a Hamiltonian $H({\bf q},{\bf p})$
\begin{equation} \label{A1}
N(E) = \frac{1}{(2\pi\hbar)^2} \int_{H({\bf q},{\bf p})\le E} d^2{\bf q}\;d^2{\bf p}.
\end{equation}
Since $H= {\bf p}^2/(2m)$, with constant zero potential energy
inside the billiard ${\cal B}$, where $m$ is the mass of the billiard
point particle, and $H$ is infinite on the boundary $\partial {\cal B}$,
we get at once
\begin{equation} \label{A2}
N(E) = \frac{2\pi {\cal A}mE}{(2\pi\hbar)^2}.
\end{equation}
Here ${\cal A}$ is the area of the billiard ${\cal B}$.
The density of levels is $\rho (E) = 1/(\Delta E) = 
dN(E)/dE = {\cal A}m/(2\pi\hbar^2)$
and thus the Heisenberg time is
\begin{equation}  \label{A3}
t_H = 2\pi\hbar \rho(E) = \frac{{\cal A} m}{\hbar}.
\end{equation}
The classical transport time is denoted by $t_T$, and in units of
the number of collisions $N_T$ can be written as 
\begin{equation} \label{A4}
t_T = \frac{\bar{l} N_T}{v} = \frac{ \bar{l}N_T}{\sqrt{2E/m}},
\end{equation}
where $\bar{l}$ is the mean free path of the billiard particle
and $v =\sqrt{2E/m}$ is its speed at the energy $E$. Thus for 
the ratio  $\alpha = t_H/t_T$ we get
\begin{equation} \label{A5}
\alpha = \frac{t_H}{t_T} = \frac{{\cal A} k}{N_T \bar{l}} 
\end{equation}
where $k= \sqrt{2mE/\hbar^2}$. Taking into account that
$\bar{l} \approx \pi {\cal A}/{\cal L}$ (this is so-called Santalo's
formula, see e.g. \cite{Santalo}), we have
\begin{equation} \label{A6}
\alpha = \frac{t_H}{t_T} = \frac{{\cal L}k}{\pi N_T},
\end{equation}
where ${\cal L}$ is the length of the perimeter $\partial {\cal B}$.
This is a general formula valid for any chaotic billiard.
In the case of the stadium billiard  with small $\ep$  we have  
${\cal L}\approx 2\pi$ and we arrive at the final estimate

\begin{equation} \label{A7}
\alpha = \frac{2k}{N_T}.
\end{equation}
Thus the condition for the occurrence of dynamical localization
$\alpha \le 1$ is now expressed in the inequality

\begin{equation} \label{A8}
k \le \frac{N_T}{2}.
\end{equation}

Of course, the definition of the classical transport time is
rather arbitrary. One definition is in terms of the diffusion time
which appears in the exponential approach of the variance of
the momentum distribution to its asymptotical value $1/3$ as
explained in detail in Ref. \cite{LozRob2018A}, where the starting
(initial) distribution is just the Dirac delta distribution $\delta(p)$.
The other possible definition of $N_T$ is by the time at
which the variance reaches certain fraction of its asymptotic value,
for which we have taken 50\%, 70\%, 80\% and 90\%. The results of
numerical calculations are shown in Table I.

\begin{figure}[H]
  \centering
  \includegraphics[width=9cm]{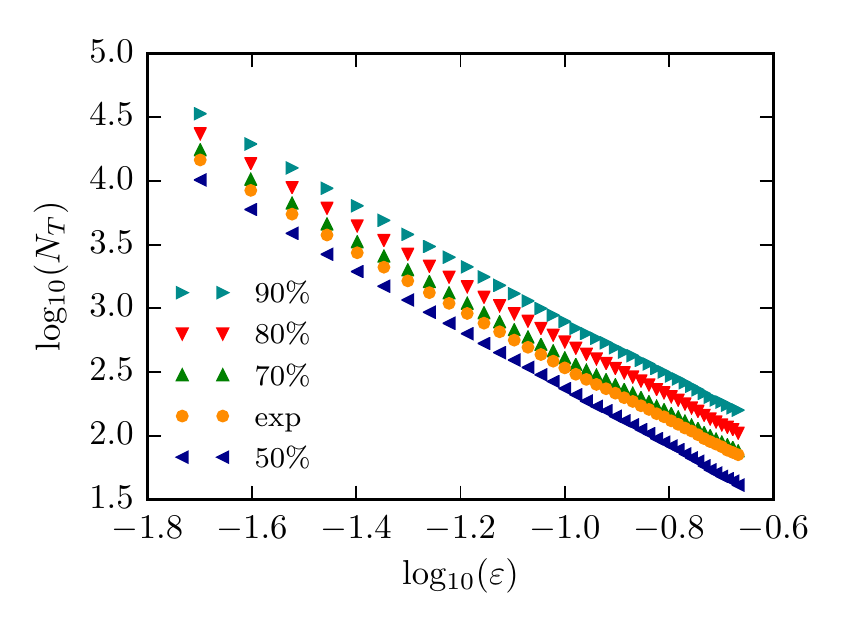}
  \caption{The transport times of Table I as functions of $\ep$ in log-log
    presentation.
    They clearly obey power laws, at smaller $\ep<0.1$ with the slope $\approx -2.33$,
    and at larger $\ep>0.1$ with the slope $\approx -2.1$., while the global fit
    (ignoring the break point at around $\ep=0.1$) gives $\approx -2.25$. For the
  precise data see Table II.}
  \label{TTimes}
\end{figure}

\begin{table}
  \center
  \begin{tabular}{ | p{1.2cm} | p{1.0cm}  | p{1.0cm} | p{1.0cm} | p{1.0cm} | p{1.0cm} | }
    \hline
    \multicolumn{6}{|c|}{Transport times}\\
    \hline
   $\ep$ &  $N_T90\%$ &   $ N_T80\%$  & $N_T70\%$ & $ N_T50\%$  &  $N_Texp$  \\ \hline
 0.0200 &  33691 &  23498 &  17516 &  10172 &    14613 \\ \hline
 0.0250 &  19486 &  13689 &  10277 &   5968 &     8429 \\ \hline
 0.0300 &  12645 &   8887 &   6691 &   3877 &     5487 \\ \hline
 0.0350 &   8758 &   6117 &   4575 &   2653 &     3765 \\ \hline
 0.0400 &   6383 &   4450 &   3322 &   1944 &     2730 \\ \hline
 0.0450 &   4900 &   3414 &   2566 &   1490 &     2104 \\ \hline
 0.0500 &   3805 &   2671 &   2001 &   1163 &     1643 \\ \hline
 0.0550 &   3061 &   2148 &   1620 &    933 &     1328 \\ \hline
 0.0600 &   2517 &   1764 &   1321 &    763 &     1094 \\ \hline
 0.0650 &   2116 &   1481 &   1094 &    633 &      912 \\ \hline
 0.0700 &   1766 &   1226 &    921 &    530 &      765 \\ \hline
 0.0750 &   1515 &   1050 &    783 &    449 &      655 \\ \hline
 0.0800 &   1305 &    909 &    679 &    393 &      563 \\ \hline
 0.0850 &   1144 &    795 &    594 &    344 &      495 \\ \hline
 0.0900 &    998 &    697 &    521 &    301 &      434 \\ \hline
 0.0950 &    885 &    618 &    463 &    267 &      385 \\ \hline
 0.1000 &    788 &    547 &    408 &    235 &      341 \\ \hline
 0.1050 &    697 &    488 &    363 &    210 &      304 \\ \hline
 0.1100 &    635 &    438 &    326 &    187 &      277 \\ \hline
 0.1150 &    581 &    402 &    298 &    170 &      253 \\ \hline
 0.1200 &    537 &    370 &    275 &    157 &      234 \\ \hline
 0.1250 &    492 &    339 &    251 &    142 &      216 \\ \hline
 0.1300 &    454 &    313 &    231 &    131 &      199 \\ \hline
 0.1350 &    425 &    290 &    215 &    122 &      186 \\ \hline
 0.1400 &    390 &    270 &    199 &    112 &      172 \\ \hline
 0.1450 &    366 &    251 &    185 &    104 &      161 \\ \hline
 0.1500 &    337 &    231 &    170 &     95 &      149 \\ \hline
 0.1550 &    317 &    218 &    160 &     89 &      141 \\ \hline
 0.1600 &    295 &    203 &    149 &     83 &      131 \\ \hline
 0.1650 &    279 &    191 &    140 &     78 &      123 \\ \hline
 0.1700 &    261 &    178 &    130 &     72 &      115 \\ \hline
 0.1750 &    245 &    166 &    121 &     67 &      109 \\ \hline
 0.1800 &    230 &    156 &    114 &     63 &      102 \\ \hline
 0.1850 &    215 &    145 &    106 &     58 &       95 \\ \hline
 0.1900 &    201 &    136 &    100 &     54 &       90 \\ \hline
 0.1950 &    191 &    129 &     94 &     51 &       86 \\ \hline
 0.2000 &    184 &    122 &     89 &     48 &       82 \\ \hline
 0.2050 &    174 &    117 &     85 &     46 &       77 \\ \hline
 0.2100 &    166 &    112 &     81 &     44 &       74 \\ \hline
 0.2150 &    159 &    105 &     76 &     41 &       71 \\ \hline
\end{tabular}\\
    \caption{The discrete transport time $N_T$ (number of collisions) 
      as function of $\ep$, in terms of criteria 90\%, 80\%, 70\% and 50\%
      of the asymptotic value of the momentum variance, and in terms of
 the  diffusion time (from the exponential law).}
\end{table}
We also show the graph of these data in Fig. \ref{TTimes}.
They clearly obey power laws with almost the same slopes, namely,
at smaller $\ep<0.1$ with the slope approximately $-2.3$,
and at larger $\ep>0.1$ with the slope approximately  $-2.1$. Thus,
they differ approximately only by an apparently $\ep$-independent
factor. The transition region around the break point $\ep\approx 0.1$
is about $0.2$ wide. The precise values of the exponents and their
estimated errors are in Table II. In the global fit (all $\ep$, ignoring the
weak break point) the exponents are indeed almost the same, approximately
$-2.25$.

\begin{table}
  \center
  \begin{tabular}{ | p{1.2cm} || p{1.0cm}  | p{1.0cm} || p{1.0cm} | p{1.0cm} || p{1.0cm} | p{1.0cm} |}
    \hline
    \multicolumn{7}{|c|}{Power law exponents for transport times}\\
    \hline
   $N_T$ & power all $\ep$ &   error  & power $\ep<0.1$ & error  &  power $\ep>0.1$ & error \\ \hline
 90\% &  -2.229 &  0.0101 &  -2.323 &  0.0079 &  -2.092 & 0.0078\\ \hline
 80\% &  -2.249 &  0.0084 &  -2.327 &  0.0076 &	 -2.142 & 0.0092\\ \hline
 70\% &  -2.263 &  0.0072 &  -2.329 &  0.0072 &	 -2.178 & 0.0096 \\ \hline
 50\% &  -2.294 &  0.0048 &  -2.333 &  0.0069 &	 -2.259 & 0.0105 \\ \hline
 exp  &  -2.211 &  0.0117 &  -2.320 &  0.0096 &	 -2.053 & 0.0078 \\ \hline
 \end{tabular}\\
 \caption{The power law exponents (with estimated errors) of $N_T$ of Fig. \ref{TTimes}
   as functions of $\ep$ in log-log presentation: The first two columns refer to the
   global fit. There is a break point at approximately $\ep\approx 0.1$
   where the slopes slightly change. The second pair of columns refers to the
   interval $\ep<0.1$, and the last two columns refer to the interval $\ep>0.1$.}
\end{table}

\section{The scaling of $\beta$ and $A$ with  $\alpha$}
\label{sec4}

\begin{figure}[H]
  \centering
  \includegraphics[width=9cm]{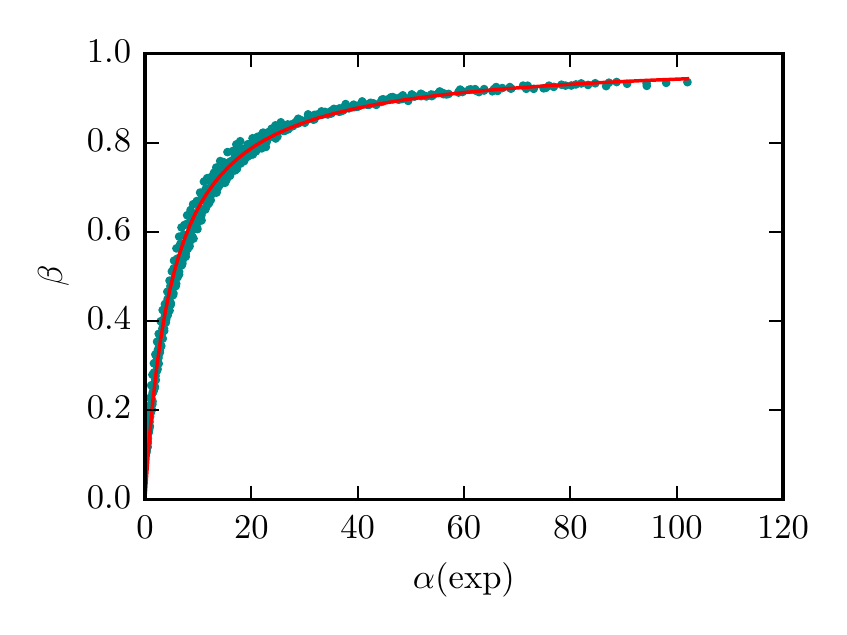}
  \caption{The level repulsion exponent $\beta$ as a function of $\al$ fitted
    by the function (\ref{BvsAExp}), based on $N_T$ from the exponential
    diffusion law. $\beta_{\infty}=0.98$ and $s=0.20$.}  
  \label{betavsalphaExp}
\end{figure}

Having established the transport times and the parameter $\al$ in
Eq. (\ref{A7}) we can now look at the dependence of the level repulsion
exponent $\beta$ on $\al$ for various definitions of $N_T$, from Table I.
For each $\beta(\epsilon_j,(k_i,k_{i+1}))$ (see Section \ref{sec2})
an associated value of $\alpha$ was
computed using Eq. (\ref{A7}) where $N_T=N_T(\epsilon_j)$ and $k=\bar{k}_i=\frac{1}{2}(k_i+k_{i+1})$.
In Fig. \ref{betavsalphaExp}, using the $N_T$ from the exponential law,
we clearly see that $\beta$ is a function of $\al$, empirically well
described by the rational function

\be  \label{BvsAExp}
\beta = \beta_{\infty} \frac{s\al}{1 +s \al}.
\ee
where the parameter $s$ depends on the definition of $N_T$, as we shall see,
and changes with the definition of $N_T$ implicit in $\al$, but the functional
form (\ref{BvsAExp}) persists. Here we find $\beta_{\infty}=0.98$ and
$s=0.20$, using the $N_T$ from the exponential diffusion law.

In the next Fig. \ref{betavsalpha} we show the dependence of $\beta$ on $\alpha$
using the definitions of the transport time in terms of the fraction of the asymptotic value of
the momentum variance. Again, the rational function (\ref{BvsAExp}) is
confirmed.

\begin{figure*}[t]
  \begin{centering}
  \includegraphics[width=1\textwidth]{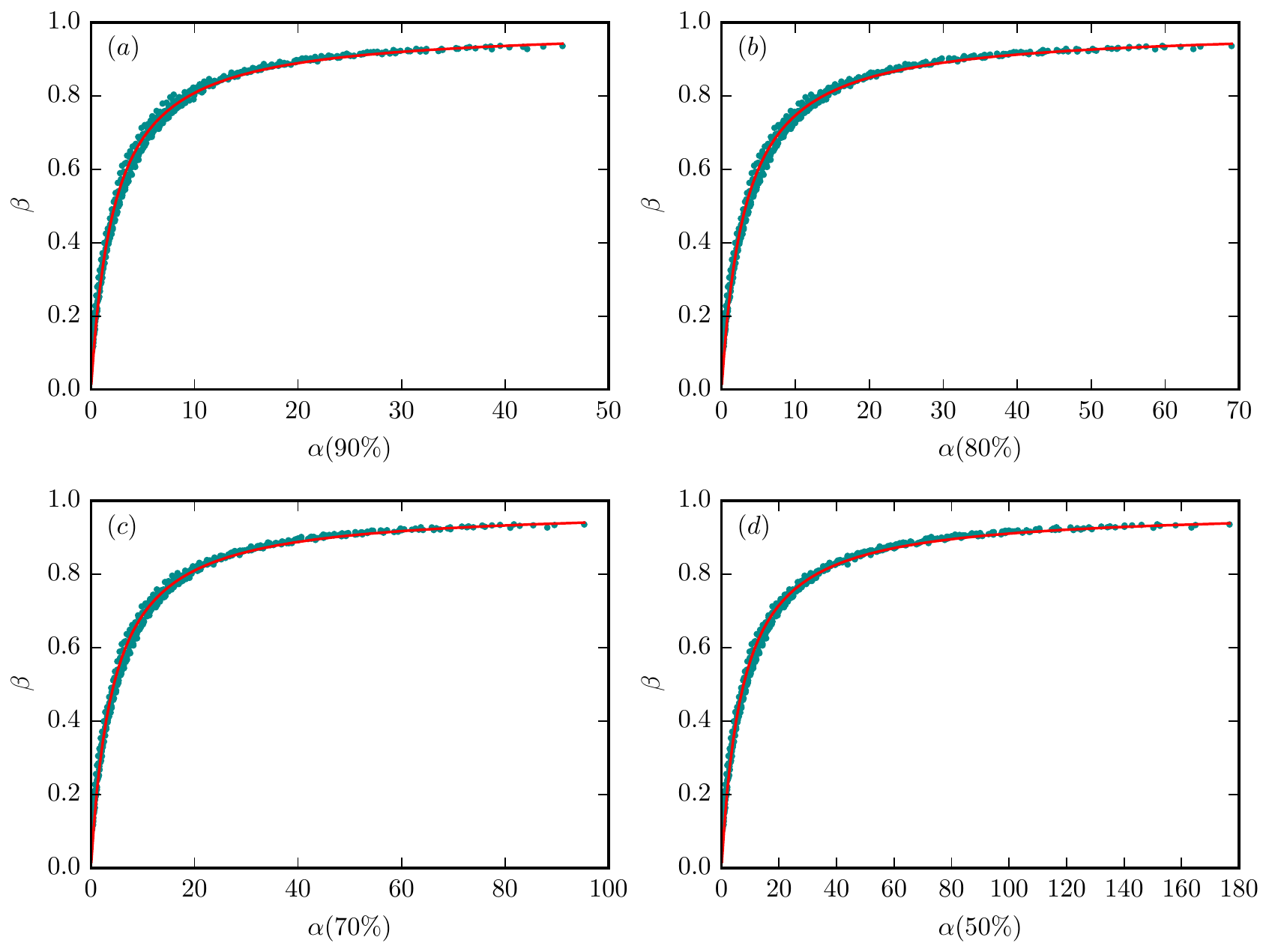}
  \par\end{centering}
  \caption{The level repulsion exponent $\beta$ as a function of $\al$ fitted
    by the function (\ref{BvsAExp}). For the transport times calculated in
    Table I, we find for $90\%$, $80\%$, $70\%$ and $50\%$ criterion
    the corresponding values ($\beta_{\infty},s$) as follows: (a) $(0.98,0.46)$, (b) $(0.98,0.32)$,
    (c) $(0.98,0.24)$, and (d) $(0.98,0.13)$.}
  \label{betavsalpha}
\end{figure*}
It should be observed that according to the empirical law of Eq. (\ref{BvsAExp}),
and as seen in both Figs. \ref{betavsalphaExp}  and \ref{betavsalpha},
the transition from complete localization $\beta=0$ to the full extendedness
(delocalization) $\beta \approx 1$ is very smooth, as it happens on the interval
of about almost two decades of $\al$, rather than being abrupt.

\begin{figure}[H]
  \centering
  \includegraphics[width=9cm]{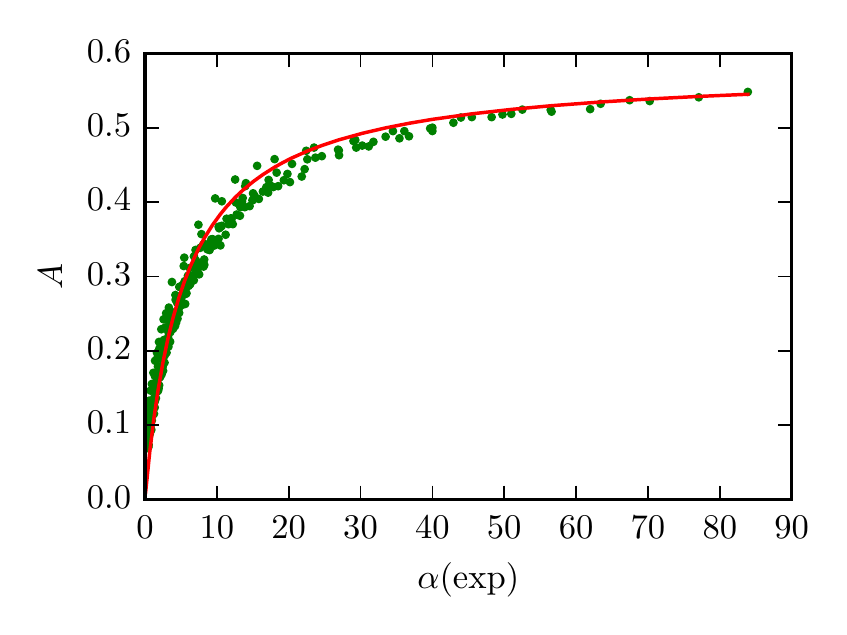}
  \caption{The entropy localization measure $A$ as a function of $\al$ fitted
  by the function (\ref{Avsalphaeq}), based on $N_T$ from the exponential
    diffusion law. $A_{\infty}=0.58$ and $s=0.19$.} 
  \label{Avsalphaexp}
\end{figure}

\begin{figure*}[t]
  \begin{centering}
    \includegraphics[width=1\textwidth]{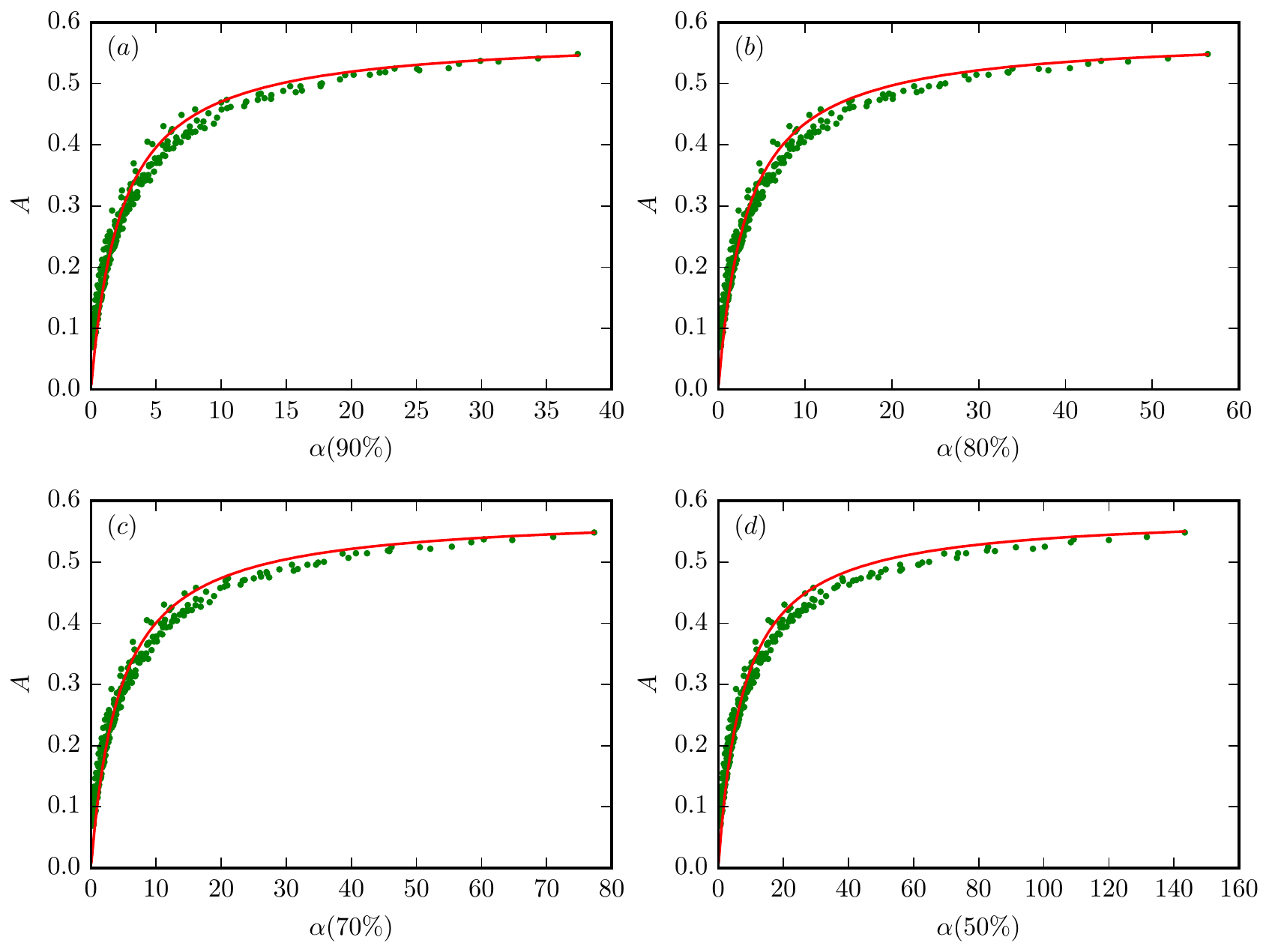}
    \par\end{centering}
  \caption{The entropy localization measure $A$ as a function of $\al$ fitted
    by the function (\ref{Avsalphaeq}). For the transport times calculated in
    Table I,  we find for $90\%$, $80\%$, $70\%$ and $50\%$ criterion 
    the corresponding values ($A_{\infty},s$) as follows: (a) $(0.58, 0.43)$, (b) $(0.58, 0.30)$,
    (c) $(0.58, 0.22)$, and (d) $(0.58, 0.13)$.}
  \label{Avsalpha}
\end{figure*}

Finally, we look at the dependence of the localization measure $A$, defined
in Eqs. (\ref{locA},\ref{entropy}), on $\al$. As we see in Fig. \ref{betaVsA},
$\beta$ is a linear function of $A$, while it is a rational function of $\al$.
Thus the entropy localization measure
$A$ also must be a rational function of $\al$, similarly as in
Eq. (\ref{BvsAExp}), namely

\be  \label{Avsalphaeq}
A = A_{\infty} \frac{s\al}{1 +s \al}.
\ee
Indeed, in Fig. \ref{Avsalphaexp} we see that this is the case.
In analogy with figures \ref{betavsalpha} we display also the dependence of $A$
on $\al$ for four various definitions of $N_T$ from Table I in Fig. \ref{Avsalpha}.

\section{Conclusions  and discussion}

Our main conclusion is that in the stadium billiard of Bunimovich \cite{Bun1979}
the spectral level repulsion exponent $\beta$ of the chaotic eigenstates
is functionally related to the localization measure, here specifically the entropy
localization measure $A$, calculated by using the Poincar\'e-Husimi functions.
Moreover, the dependence is linear, as in the
quantum kicked rotator, but different from the case of a mixed type billiard
studied recently by Batisti\'c and Robnik \cite{BatRob2013A,BatRob2013B}, where the
high-lying localized chaotic eigenstates have been analyzed after the separation of
regular and chaotic eigenstates.

Furthermore, we have shown that $\beta$ is a rational function of the
major control parameter $\al$, which is the ratio of the Heisenberg time
and the classical transport time. The definition of the classical transport
time is to some extent arbitrary, but we have shown that the various
definitions do not change the shape of the dependence on $\ep$, but instead affect
only the prefactor. As a cosequence of that the dependence is always
a rational function. The transition from complete localization $\beta=0$
to the complete extendedness (delocalization) $\beta\approx 1$
takes place very smoothly, over about two decades of the parameter $\al$.

Thus we have again demonstrated by numerical
calculation that the fractional power law level repulsion
with the exponent $\beta \in [0,1]$ is manifested in localized chaotic eigenstates.
Our empirical findings call for theoretical explanation, which is
a long standing open problem even for the main paradigm of quantum chaos,
the quantum kicked rotator studied extensively over the decades
\cite{Izr1990}.

Further theoretical work is in progress. Beyond the billiard systems,
there are many important applications in various physical systems,
like e.g. in hydrogen atom in strong magnetic field \cite{Rob1981,Rob1982,HRW1989,
WF1989,RWHG1994}, which is a paradigm of stationary quantum chaos,
or e.g. in microwave resonators, the experiments introduced by St\"ockmann
around 1990 and intensely further developed since then \cite{Stoe}.

\section{Acknowledgement}

This work was supported by the Slovenian Research Agency (ARRS) under
the grant J1-9112.


\begin{thebibliography}{47}%
\makeatletter
\providecommand \@ifxundefined [1]{%
 \@ifx{#1\undefined}
}%
\providecommand \@ifnum [1]{%
 \ifnum #1\expandafter \@firstoftwo
 \else \expandafter \@secondoftwo
 \fi
}%
\providecommand \@ifx [1]{%
 \ifx #1\expandafter \@firstoftwo
 \else \expandafter \@secondoftwo
 \fi
}%
\providecommand \natexlab [1]{#1}%
\providecommand \enquote  [1]{``#1''}%
\providecommand \bibnamefont  [1]{#1}%
\providecommand \bibfnamefont [1]{#1}%
\providecommand \citenamefont [1]{#1}%
\providecommand \href@noop [0]{\@secondoftwo}%
\providecommand \href [0]{\begingroup \@sanitize@url \@href}%
\providecommand \@href[1]{\@@startlink{#1}\@@href}%
\providecommand \@@href[1]{\endgroup#1\@@endlink}%
\providecommand \@sanitize@url [0]{\catcode `\\12\catcode `\$12\catcode
  `\&12\catcode `\#12\catcode `\^12\catcode `\_12\catcode `\%12\relax}%
\providecommand \@@startlink[1]{}%
\providecommand \@@endlink[0]{}%
\providecommand \url  [0]{\begingroup\@sanitize@url \@url }%
\providecommand \@url [1]{\endgroup\@href {#1}{\urlprefix }}%
\providecommand \urlprefix  [0]{URL }%
\providecommand \Eprint [0]{\href }%
\providecommand \doibase [0]{http://dx.doi.org/}%
\providecommand \selectlanguage [0]{\@gobble}%
\providecommand \bibinfo  [0]{\@secondoftwo}%
\providecommand \bibfield  [0]{\@secondoftwo}%
\providecommand \translation [1]{[#1]}%
\providecommand \BibitemOpen [0]{}%
\providecommand \bibitemStop [0]{}%
\providecommand \bibitemNoStop [0]{.\EOS\space}%
\providecommand \EOS [0]{\spacefactor3000\relax}%
\providecommand \BibitemShut  [1]{\csname bibitem#1\endcsname}%
\let\auto@bib@innerbib\@empty
\bibitem [{\citenamefont {St\"ockmann}(1999)}]{Stoe}%
  \BibitemOpen
  \bibfield  {author} {\bibinfo {author} {\bibfnamefont {H.-J.}\ \bibnamefont
  {St\"ockmann}},\ }\href@noop {} {\emph {\bibinfo {title} {Quantum Chaos - An
  Introduction}}}\ (\bibinfo  {publisher} {Cambridge: Cambridge University
  Press},\ \bibinfo {year} {1999})\BibitemShut {NoStop}%
\bibitem [{\citenamefont {Haake}(2001)}]{Haake}%
  \BibitemOpen
  \bibfield  {author} {\bibinfo {author} {\bibfnamefont {F.}~\bibnamefont
  {Haake}},\ }\href@noop {} {\emph {\bibinfo {title} {Quantum Signatures of
  Chaos}}}\ (\bibinfo  {publisher} {Berlin: Springer},\ \bibinfo {year}
  {2001})\BibitemShut {NoStop}%
\bibitem [{\citenamefont {Casati}\ \emph {et~al.}(1979)\citenamefont {Casati},
  \citenamefont {Chirikov}, \citenamefont {Izrailev},\ and\ \citenamefont
  {Ford}}]{Cas1979}%
  \BibitemOpen
  \bibfield  {author} {\bibinfo {author} {\bibfnamefont {G.}~\bibnamefont
  {Casati}}, \bibinfo {author} {\bibfnamefont {B.~V.}\ \bibnamefont
  {Chirikov}}, \bibinfo {author} {\bibfnamefont {F.~M.}\ \bibnamefont
  {Izrailev}}, \ and\ \bibinfo {author} {\bibfnamefont {J.}~\bibnamefont
  {Ford}},\ }\href@noop {} {\bibfield  {journal} {\bibinfo  {journal} {Lecture
  Notes in Physics}\ }\textbf {\bibinfo {volume} {93}},\ \bibinfo {pages} {334}
  (\bibinfo {year} {1979})}\BibitemShut {NoStop}%
\bibitem [{\citenamefont {Chirikov}\ \emph {et~al.}(1981)\citenamefont
  {Chirikov}, \citenamefont {Izrailev},\ and\ \citenamefont
  {Shepelyansky}}]{Chi1981}%
  \BibitemOpen
  \bibfield  {author} {\bibinfo {author} {\bibfnamefont {B.~V.}\ \bibnamefont
  {Chirikov}}, \bibinfo {author} {\bibfnamefont {F.~M.}\ \bibnamefont
  {Izrailev}}, \ and\ \bibinfo {author} {\bibfnamefont {D.~L.}\ \bibnamefont
  {Shepelyansky}},\ }\href@noop {} {\bibfield  {journal} {\bibinfo  {journal}
  {Sov. Sci. Rev. C}\ }\textbf {\bibinfo {volume} {2}},\ \bibinfo {pages} {209}
  (\bibinfo {year} {1981})}\BibitemShut {NoStop}%
\bibitem [{\citenamefont {Chirikov}\ \emph {et~al.}(1988)\citenamefont
  {Chirikov}, \citenamefont {Izrailev},\ and\ \citenamefont
  {Shepelyansky}}]{Chi1988}%
  \BibitemOpen
  \bibfield  {author} {\bibinfo {author} {\bibfnamefont {B.~V.}\ \bibnamefont
  {Chirikov}}, \bibinfo {author} {\bibfnamefont {F.~M.}\ \bibnamefont
  {Izrailev}}, \ and\ \bibinfo {author} {\bibfnamefont {D.~L.}\ \bibnamefont
  {Shepelyansky}},\ }\href@noop {} {\bibfield  {journal} {\bibinfo  {journal}
  {Physica D}\ }\textbf {\bibinfo {volume} {33}},\ \bibinfo {pages} {77}
  (\bibinfo {year} {1988})}\BibitemShut {NoStop}%
\bibitem [{\citenamefont {Izrailev}(1990)}]{Izr1990}%
  \BibitemOpen
  \bibfield  {author} {\bibinfo {author} {\bibfnamefont {F.~M.}\ \bibnamefont
  {Izrailev}},\ }\href@noop {} {\bibfield  {journal} {\bibinfo  {journal}
  {Phys. Rep.}\ }\textbf {\bibinfo {volume} {196}},\ \bibinfo {pages} {299}
  (\bibinfo {year} {1990})}\BibitemShut {NoStop}%
\bibitem [{\citenamefont {Izrailev}(1988)}]{Izr1988}%
  \BibitemOpen
  \bibfield  {author} {\bibinfo {author} {\bibfnamefont {F.~M.}\ \bibnamefont
  {Izrailev}},\ }\href@noop {} {\bibfield  {journal} {\bibinfo  {journal}
  {Phys. Lett. A}\ }\textbf {\bibinfo {volume} {134}},\ \bibinfo {pages} {13}
  (\bibinfo {year} {1988})}\BibitemShut {NoStop}%
\bibitem [{\citenamefont {Izrailev}(1989)}]{Izr1989}%
  \BibitemOpen
  \bibfield  {author} {\bibinfo {author} {\bibfnamefont {F.~M.}\ \bibnamefont
  {Izrailev}},\ }\href@noop {} {\bibfield  {journal} {\bibinfo  {journal} {J.
  Phys. A: Math. Gen.}\ }\textbf {\bibinfo {volume} {22}},\ \bibinfo {pages}
  {865} (\bibinfo {year} {1989})}\BibitemShut {NoStop}%
\bibitem [{\citenamefont {Fishman}\ \emph {et~al.}(1982)\citenamefont
  {Fishman}, \citenamefont {Grempel},\ and\ \citenamefont {Prange}}]{FGP1982}%
  \BibitemOpen
  \bibfield  {author} {\bibinfo {author} {\bibfnamefont {S.}~\bibnamefont
  {Fishman}}, \bibinfo {author} {\bibfnamefont {D.~R.}\ \bibnamefont
  {Grempel}}, \ and\ \bibinfo {author} {\bibfnamefont {R.~E.}\ \bibnamefont
  {Prange}},\ }\href@noop {} {\bibfield  {journal} {\bibinfo  {journal} {Phys.
  Rev. Lett.}\ }\textbf {\bibinfo {volume} {49}},\ \bibinfo {pages} {509}
  (\bibinfo {year} {1982})}\BibitemShut {NoStop}%
\bibitem [{\citenamefont {Prosen}(2000)}]{Pro2000}%
  \BibitemOpen
  \bibfield  {author} {\bibinfo {author} {\bibfnamefont {T.}~\bibnamefont
  {Prosen}},\ }\href@noop {} {\emph {\bibinfo {title} {in Proc. of the Int.
  School in Phys. "Enrico Fermi", Course CXLIII, Eds. G. Casati and U.
  Smilansky}}}\ (\bibinfo  {publisher} {Amsterdam: IOS Press},\ \bibinfo {year}
  {2000})\BibitemShut {NoStop}%
\bibitem [{\citenamefont {Batisti\'c}\ and\ \citenamefont
  {Robnik}(2013{\natexlab{a}})}]{BatRob2013A}%
  \BibitemOpen
  \bibfield  {author} {\bibinfo {author} {\bibfnamefont {B.}~\bibnamefont
  {Batisti\'c}}\ and\ \bibinfo {author} {\bibfnamefont {M.}~\bibnamefont
  {Robnik}},\ }\href@noop {} {\bibfield  {journal} {\bibinfo  {journal} {Phys.
  Rev. E}\ }\textbf {\bibinfo {volume} {88}},\ \bibinfo {pages} {052913}
  (\bibinfo {year} {2013}{\natexlab{a}})}\BibitemShut {NoStop}%
\bibitem [{\citenamefont {Robnik}(1983)}]{Rob1983}%
  \BibitemOpen
  \bibfield  {author} {\bibinfo {author} {\bibfnamefont {M.}~\bibnamefont
  {Robnik}},\ }\href@noop {} {\bibfield  {journal} {\bibinfo  {journal} {J.
  Phys. A: Math. Gen.}\ }\textbf {\bibinfo {volume} {16}},\ \bibinfo {pages}
  {3971} (\bibinfo {year} {1983})}\BibitemShut {NoStop}%
\bibitem [{\citenamefont {Robnik}(1984)}]{Rob1984}%
  \BibitemOpen
  \bibfield  {author} {\bibinfo {author} {\bibfnamefont {M.}~\bibnamefont
  {Robnik}},\ }\href@noop {} {\bibfield  {journal} {\bibinfo  {journal} {J.
  Phys. A: Math. Gen.}\ }\textbf {\bibinfo {volume} {17}},\ \bibinfo {pages}
  {1049} (\bibinfo {year} {1984})}\BibitemShut {NoStop}%
\bibitem [{\citenamefont {Batisti\'c}\ and\ \citenamefont
  {Robnik}(2013{\natexlab{b}})}]{BatRob2013B}%
  \BibitemOpen
  \bibfield  {author} {\bibinfo {author} {\bibfnamefont {B.}~\bibnamefont
  {Batisti\'c}}\ and\ \bibinfo {author} {\bibfnamefont {M.}~\bibnamefont
  {Robnik}},\ }\href@noop {} {\bibfield  {journal} {\bibinfo  {journal} {J.
  Phys. A: Math. Theor.}\ }\textbf {\bibinfo {volume} {46}},\ \bibinfo {pages}
  {315102} (\bibinfo {year} {2013}{\natexlab{b}})}\BibitemShut {NoStop}%
\bibitem [{\citenamefont {Bunimovich}(1979)}]{Bun1979}%
  \BibitemOpen
  \bibfield  {author} {\bibinfo {author} {\bibfnamefont {L.}~\bibnamefont
  {Bunimovich}},\ }\href@noop {} {\bibfield  {journal} {\bibinfo  {journal}
  {Commun. Math. Phys.}\ }\textbf {\bibinfo {volume} {65}},\ \bibinfo {pages}
  {295} (\bibinfo {year} {1979})}\BibitemShut {NoStop}%
\bibitem [{\citenamefont {Borgonovi}\ \emph {et~al.}(1996)\citenamefont
  {Borgonovi}, \citenamefont {Casati},\ and\ \citenamefont {Li}}]{BCL1996}%
  \BibitemOpen
  \bibfield  {author} {\bibinfo {author} {\bibfnamefont {F.}~\bibnamefont
  {Borgonovi}}, \bibinfo {author} {\bibfnamefont {G.}~\bibnamefont {Casati}}, \
  and\ \bibinfo {author} {\bibfnamefont {B.}~\bibnamefont {Li}},\ }\href@noop
  {} {\bibfield  {journal} {\bibinfo  {journal} {Phys. Rev. Lett.}\ }\textbf
  {\bibinfo {volume} {77}},\ \bibinfo {pages} {4744} (\bibinfo {year}
  {1996})}\BibitemShut {NoStop}%
\bibitem [{\citenamefont {\v{C}. Lozej}\ and\ \citenamefont
  {Robnik}(2018)}]{LozRob2018A}%
  \BibitemOpen
  \bibfield  {author} {\bibinfo {author} {\bibnamefont {\v{C}. Lozej}}\ and\
  \bibinfo {author} {\bibfnamefont {M.}~\bibnamefont {Robnik}},\ }\href@noop {}
  {\bibfield  {journal} {\bibinfo  {journal} {Phys. Rev. E}\ }\textbf {\bibinfo
  {volume} {97}},\ \bibinfo {pages} {012206} (\bibinfo {year}
  {2018})}\BibitemShut {NoStop}%
\bibitem [{\citenamefont {Mehta}(1991)}]{Mehta}%
  \BibitemOpen
  \bibfield  {author} {\bibinfo {author} {\bibfnamefont {M.~L.}\ \bibnamefont
  {Mehta}},\ }\href@noop {} {\emph {\bibinfo {title} {Random Matrices}}}\
  (\bibinfo  {publisher} {Boston: Academic Press},\ \bibinfo {year}
  {1991})\BibitemShut {NoStop}%
\bibitem [{\citenamefont {Guhr}\ \emph {et~al.}(1998)\citenamefont {Guhr},
  \citenamefont {M\"uller-Groeling},\ and\ \citenamefont
  {Weidenm\"uller}}]{GMW}%
  \BibitemOpen
  \bibfield  {author} {\bibinfo {author} {\bibfnamefont {T.}~\bibnamefont
  {Guhr}}, \bibinfo {author} {\bibfnamefont {A.}~\bibnamefont
  {M\"uller-Groeling}}, \ and\ \bibinfo {author} {\bibfnamefont
  {H.}~\bibnamefont {Weidenm\"uller}},\ }\href@noop {} {\bibfield  {journal}
  {\bibinfo  {journal} {Phys. Rep.}\ }\textbf {\bibinfo {volume} {299}},\
  \bibinfo {pages} {4} (\bibinfo {year} {1998})}\BibitemShut {NoStop}%
\bibitem [{\citenamefont {Robnik}(1998)}]{Rob1998}%
  \BibitemOpen
  \bibfield  {author} {\bibinfo {author} {\bibfnamefont {M.}~\bibnamefont
  {Robnik}},\ }\href@noop {} {\bibfield  {journal} {\bibinfo  {journal} {Nonl.
  Phen. in Compl. Syst. (Minsk)}\ }\textbf {\bibinfo {volume} {1}},\ \bibinfo
  {pages} {1} (\bibinfo {year} {1998})}\BibitemShut {NoStop}%
\bibitem [{\citenamefont {Percival}(1973)}]{Percival1973}%
  \BibitemOpen
  \bibfield  {author} {\bibinfo {author} {\bibfnamefont {I.~C.}\ \bibnamefont
  {Percival}},\ }\href@noop {} {\bibfield  {journal} {\bibinfo  {journal} {J.
  Phys B: At. Mol. Phys.}\ }\textbf {\bibinfo {volume} {6}},\ \bibinfo {pages}
  {L229} (\bibinfo {year} {1973})}\BibitemShut {NoStop}%
\bibitem [{\citenamefont {Berry}\ and\ \citenamefont
  {Robnik}(1984)}]{BerRob1984}%
  \BibitemOpen
  \bibfield  {author} {\bibinfo {author} {\bibfnamefont {M.~V.}\ \bibnamefont
  {Berry}}\ and\ \bibinfo {author} {\bibfnamefont {M.}~\bibnamefont {Robnik}},\
  }\href@noop {} {\bibfield  {journal} {\bibinfo  {journal} {J. Phys. A: Math.
  Gen.}\ }\textbf {\bibinfo {volume} {17}},\ \bibinfo {pages} {2413} (\bibinfo
  {year} {1984})}\BibitemShut {NoStop}%
\bibitem [{\citenamefont {Batisti\'c}\ and\ \citenamefont
  {Robnik}(2010)}]{BatRob2010}%
  \BibitemOpen
  \bibfield  {author} {\bibinfo {author} {\bibfnamefont {B.}~\bibnamefont
  {Batisti\'c}}\ and\ \bibinfo {author} {\bibfnamefont {M.}~\bibnamefont
  {Robnik}},\ }\href@noop {} {\bibfield  {journal} {\bibinfo  {journal} {J.
  Phys. A: Math. Theor.}\ }\textbf {\bibinfo {volume} {43}},\ \bibinfo {pages}
  {215101} (\bibinfo {year} {2010})}\BibitemShut {NoStop}%
\bibitem [{\citenamefont {Casati}\ \emph {et~al.}(1980)\citenamefont {Casati},
  \citenamefont {Valz-Gris},\ and\ \citenamefont {Guarneri}}]{Cas1980}%
  \BibitemOpen
  \bibfield  {author} {\bibinfo {author} {\bibfnamefont {G.}~\bibnamefont
  {Casati}}, \bibinfo {author} {\bibfnamefont {F.}~\bibnamefont {Valz-Gris}}, \
  and\ \bibinfo {author} {\bibfnamefont {I.}~\bibnamefont {Guarneri}},\
  }\href@noop {} {\bibfield  {journal} {\bibinfo  {journal} {Lett. Nuovo
  Cimento}\ }\textbf {\bibinfo {volume} {28}},\ \bibinfo {pages} {279}
  (\bibinfo {year} {1980})}\BibitemShut {NoStop}%
\bibitem [{\citenamefont {Bohigas}\ \emph {et~al.}(1984)\citenamefont
  {Bohigas}, \citenamefont {Giannoni},\ and\ \citenamefont {Schmit}}]{BGS1984}%
  \BibitemOpen
  \bibfield  {author} {\bibinfo {author} {\bibfnamefont {O.}~\bibnamefont
  {Bohigas}}, \bibinfo {author} {\bibfnamefont {M.~J.}\ \bibnamefont
  {Giannoni}}, \ and\ \bibinfo {author} {\bibfnamefont {C.}~\bibnamefont
  {Schmit}},\ }\href@noop {} {\bibfield  {journal} {\bibinfo  {journal} {Phys.
  Rev. Lett.}\ }\textbf {\bibinfo {volume} {52}},\ \bibinfo {pages} {1}
  (\bibinfo {year} {1984})}\BibitemShut {NoStop}%
\bibitem [{\citenamefont {Sieber}\ and\ \citenamefont
  {Richter}(2001)}]{Sieber}%
  \BibitemOpen
  \bibfield  {author} {\bibinfo {author} {\bibfnamefont {M.}~\bibnamefont
  {Sieber}}\ and\ \bibinfo {author} {\bibfnamefont {K.}~\bibnamefont
  {Richter}},\ }\href@noop {} {\bibfield  {journal} {\bibinfo  {journal} {Phys.
  Scr.}\ }\textbf {\bibinfo {volume} {T90}},\ \bibinfo {pages} {128} (\bibinfo
  {year} {2001})}\BibitemShut {NoStop}%
\bibitem [{\citenamefont {M\"uller}\ \emph {et~al.}(2004)\citenamefont
  {M\"uller}, \citenamefont {Heusler}, \citenamefont {Braun}, \citenamefont
  {Haake},\ and\ \citenamefont {Altland}}]{Mueller1}%
  \BibitemOpen
  \bibfield  {author} {\bibinfo {author} {\bibfnamefont {S.}~\bibnamefont
  {M\"uller}}, \bibinfo {author} {\bibfnamefont {S.}~\bibnamefont {Heusler}},
  \bibinfo {author} {\bibfnamefont {P.}~\bibnamefont {Braun}}, \bibinfo
  {author} {\bibfnamefont {F.}~\bibnamefont {Haake}}, \ and\ \bibinfo {author}
  {\bibfnamefont {A.}~\bibnamefont {Altland}},\ }\href@noop {} {\bibfield
  {journal} {\bibinfo  {journal} {Phys. Rev. Lett.}\ }\textbf {\bibinfo
  {volume} {93}},\ \bibinfo {pages} {014103} (\bibinfo {year}
  {2004})}\BibitemShut {NoStop}%
\bibitem [{\citenamefont {Heusler}\ \emph {et~al.}(2004)\citenamefont
  {Heusler}, \citenamefont {M\"uller}, \citenamefont {Braun},\ and\
  \citenamefont {Haake}}]{Mueller2}%
  \BibitemOpen
  \bibfield  {author} {\bibinfo {author} {\bibfnamefont {S.}~\bibnamefont
  {Heusler}}, \bibinfo {author} {\bibfnamefont {S.}~\bibnamefont {M\"uller}},
  \bibinfo {author} {\bibfnamefont {P.}~\bibnamefont {Braun}}, \ and\ \bibinfo
  {author} {\bibfnamefont {F.}~\bibnamefont {Haake}},\ }\href@noop {}
  {\bibfield  {journal} {\bibinfo  {journal} {J. Phys.A: Math. Gen.}\ }\textbf
  {\bibinfo {volume} {37}},\ \bibinfo {pages} {L31} (\bibinfo {year}
  {2004})}\BibitemShut {NoStop}%
\bibitem [{\citenamefont {M\"uller}\ \emph {et~al.}(2005)\citenamefont
  {M\"uller}, \citenamefont {Heusler}, \citenamefont {Braun}, \citenamefont
  {Haake},\ and\ \citenamefont {Altland}}]{Mueller3}%
  \BibitemOpen
  \bibfield  {author} {\bibinfo {author} {\bibfnamefont {S.}~\bibnamefont
  {M\"uller}}, \bibinfo {author} {\bibfnamefont {S.}~\bibnamefont {Heusler}},
  \bibinfo {author} {\bibfnamefont {P.}~\bibnamefont {Braun}}, \bibinfo
  {author} {\bibfnamefont {F.}~\bibnamefont {Haake}}, \ and\ \bibinfo {author}
  {\bibfnamefont {A.}~\bibnamefont {Altland}},\ }\href@noop {} {\bibfield
  {journal} {\bibinfo  {journal} {Phys. Rev. E}\ }\textbf {\bibinfo {volume}
  {72}},\ \bibinfo {pages} {046207} (\bibinfo {year} {2005})}\BibitemShut
  {NoStop}%
\bibitem [{\citenamefont {M\"uller}\ \emph {et~al.}(2009)\citenamefont
  {M\"uller}, \citenamefont {Heusler}, \citenamefont {Altland}, \citenamefont
  {Braun},\ and\ \citenamefont {Haake}}]{Mueller4}%
  \BibitemOpen
  \bibfield  {author} {\bibinfo {author} {\bibfnamefont {S.}~\bibnamefont
  {M\"uller}}, \bibinfo {author} {\bibfnamefont {S.}~\bibnamefont {Heusler}},
  \bibinfo {author} {\bibfnamefont {A.}~\bibnamefont {Altland}}, \bibinfo
  {author} {\bibfnamefont {P.}~\bibnamefont {Braun}}, \ and\ \bibinfo {author}
  {\bibfnamefont {F.}~\bibnamefont {Haake}},\ }\href@noop {} {\bibfield
  {journal} {\bibinfo  {journal} {New J. of Phys.}\ }\textbf {\bibinfo {volume}
  {11}},\ \bibinfo {pages} {103025} (\bibinfo {year} {2009})}\BibitemShut
  {NoStop}%
\bibitem [{\citenamefont {Gutzwiller}(1980)}]{Gutzwiller1980}%
  \BibitemOpen
  \bibfield  {author} {\bibinfo {author} {\bibfnamefont {M.~C.}\ \bibnamefont
  {Gutzwiller}},\ }\href@noop {} {\bibfield  {journal} {\bibinfo  {journal}
  {Phys. Rev. Lett.}\ }\textbf {\bibinfo {volume} {45}},\ \bibinfo {pages}
  {150} (\bibinfo {year} {1980})}\BibitemShut {NoStop}%
\bibitem [{\citenamefont {Berry}(1985)}]{Berry1985}%
  \BibitemOpen
  \bibfield  {author} {\bibinfo {author} {\bibfnamefont {M.~V.}\ \bibnamefont
  {Berry}},\ }\href@noop {} {\bibfield  {journal} {\bibinfo  {journal} {Proc.
  Roy. Soc. Lond. A}\ }\textbf {\bibinfo {volume} {400}},\ \bibinfo {pages}
  {229} (\bibinfo {year} {1985})}\BibitemShut {NoStop}%
\bibitem [{\citenamefont {Brody}(1973)}]{Bro1973}%
  \BibitemOpen
  \bibfield  {author} {\bibinfo {author} {\bibfnamefont {T.~A.}\ \bibnamefont
  {Brody}},\ }\href@noop {} {\bibfield  {journal} {\bibinfo  {journal} {Lett.
  Nuovo Cimento}\ }\textbf {\bibinfo {volume} {7}},\ \bibinfo {pages} {482}
  (\bibinfo {year} {1973})}\BibitemShut {NoStop}%
\bibitem [{\citenamefont {Brody}\ \emph {et~al.}(1981)\citenamefont {Brody},
  \citenamefont {Flores}, \citenamefont {French}, \citenamefont {Mello},
  \citenamefont {Pandey},\ and\ \citenamefont {Wong}}]{Bro1981}%
  \BibitemOpen
  \bibfield  {author} {\bibinfo {author} {\bibfnamefont {T.~A.}\ \bibnamefont
  {Brody}}, \bibinfo {author} {\bibfnamefont {J.}~\bibnamefont {Flores}},
  \bibinfo {author} {\bibfnamefont {J.~B.}\ \bibnamefont {French}}, \bibinfo
  {author} {\bibfnamefont {P.~A.}\ \bibnamefont {Mello}}, \bibinfo {author}
  {\bibfnamefont {A.}~\bibnamefont {Pandey}}, \ and\ \bibinfo {author}
  {\bibfnamefont {S.~S.~M.}\ \bibnamefont {Wong}},\ }\href@noop {} {\bibfield
  {journal} {\bibinfo  {journal} {Rev. Mod. Phys.}\ }\textbf {\bibinfo {volume}
  {53}},\ \bibinfo {pages} {385} (\bibinfo {year} {1981})}\BibitemShut
  {NoStop}%
\bibitem [{\citenamefont {Manos}\ and\ \citenamefont
  {Robnik}(2013)}]{ManRob2013}%
  \BibitemOpen
  \bibfield  {author} {\bibinfo {author} {\bibfnamefont {T.}~\bibnamefont
  {Manos}}\ and\ \bibinfo {author} {\bibfnamefont {M.}~\bibnamefont {Robnik}},\
  }\href@noop {} {\bibfield  {journal} {\bibinfo  {journal} {Phys. Rev. E}\
  }\textbf {\bibinfo {volume} {87}},\ \bibinfo {pages} {062905} (\bibinfo
  {year} {2013})}\BibitemShut {NoStop}%
\bibitem [{\citenamefont {Batisti{\'c}}\ \emph {et~al.}(2013)\citenamefont
  {Batisti{\'c}}, \citenamefont {Manos},\ and\ \citenamefont
  {Robnik}}]{BatManRob2013}%
  \BibitemOpen
  \bibfield  {author} {\bibinfo {author} {\bibfnamefont {B.}~\bibnamefont
  {Batisti{\'c}}}, \bibinfo {author} {\bibfnamefont {T.}~\bibnamefont {Manos}},
  \ and\ \bibinfo {author} {\bibfnamefont {M.}~\bibnamefont {Robnik}},\
  }\href@noop {} {\bibfield  {journal} {\bibinfo  {journal} {EPL}\ }\textbf
  {\bibinfo {volume} {102}},\ \bibinfo {pages} {50008} (\bibinfo {year}
  {2013})}\BibitemShut {NoStop}%
\bibitem [{\citenamefont {Wigner}(1932)}]{Wig1932}%
  \BibitemOpen
  \bibfield  {author} {\bibinfo {author} {\bibfnamefont {E.}~\bibnamefont
  {Wigner}},\ }\href@noop {} {\bibfield  {journal} {\bibinfo  {journal} {Phys.
  Rev.}\ }\textbf {\bibinfo {volume} {40}},\ \bibinfo {pages} {749} (\bibinfo
  {year} {1932})}\BibitemShut {NoStop}%
\bibitem [{\citenamefont {Husimi}(1940)}]{Hus1940}%
  \BibitemOpen
  \bibfield  {author} {\bibinfo {author} {\bibfnamefont {K.}~\bibnamefont
  {Husimi}},\ }\href@noop {} {\bibfield  {journal} {\bibinfo  {journal} {Proc.
  Phys. Math. Soc. Jpn.}\ }\textbf {\bibinfo {volume} {22}},\ \bibinfo {pages}
  {264} (\bibinfo {year} {1940})}\BibitemShut {NoStop}%
\bibitem [{\citenamefont {Tualle}\ and\ \citenamefont {Voros}(1995)}]{TV1995}%
  \BibitemOpen
  \bibfield  {author} {\bibinfo {author} {\bibfnamefont {J.}~\bibnamefont
  {Tualle}}\ and\ \bibinfo {author} {\bibfnamefont {A.}~\bibnamefont {Voros}},\
  }\href@noop {} {\bibfield  {journal} {\bibinfo  {journal} {Chaos Solitons
  Fractals}\ }\textbf {\bibinfo {volume} {5}},\ \bibinfo {pages} {1085}
  (\bibinfo {year} {1995})}\BibitemShut {NoStop}%
\bibitem [{\citenamefont {B\"acker}\ \emph {et~al.}(2004)\citenamefont
  {B\"acker}, \citenamefont {F\"urstberger},\ and\ \citenamefont
  {Schubert}}]{Baecker2004}%
  \BibitemOpen
  \bibfield  {author} {\bibinfo {author} {\bibfnamefont {A.}~\bibnamefont
  {B\"acker}}, \bibinfo {author} {\bibfnamefont {S.}~\bibnamefont
  {F\"urstberger}}, \ and\ \bibinfo {author} {\bibfnamefont {R.}~\bibnamefont
  {Schubert}},\ }\href@noop {} {\bibfield  {journal} {\bibinfo  {journal}
  {Phys. Rev. E}\ }\textbf {\bibinfo {volume} {70}},\ \bibinfo {pages} {036204}
  (\bibinfo {year} {2004})}\BibitemShut {NoStop}%
\bibitem [{\citenamefont {Manos}\ and\ \citenamefont
  {Robnik}(2015)}]{ManRob2015}%
  \BibitemOpen
  \bibfield  {author} {\bibinfo {author} {\bibfnamefont {T.}~\bibnamefont
  {Manos}}\ and\ \bibinfo {author} {\bibfnamefont {M.}~\bibnamefont {Robnik}},\
  }\href@noop {} {\bibfield  {journal} {\bibinfo  {journal} {Phys. Rev. E}\
  }\textbf {\bibinfo {volume} {91}},\ \bibinfo {pages} {042904} (\bibinfo
  {year} {2015})}\BibitemShut {NoStop}%
\bibitem [{\citenamefont {Santal\'o}\ and\ \citenamefont
  {Kac}(2004)}]{Santalo}%
  \BibitemOpen
  \bibfield  {author} {\bibinfo {author} {\bibfnamefont {L.~A.}\ \bibnamefont
  {Santal\'o}}\ and\ \bibinfo {author} {\bibfnamefont {M.}~\bibnamefont
  {Kac}},\ }\href@noop {} {\emph {\bibinfo {title} {Integral geometry and
  geometric probability}}},\ Cambridge mathematical library\ (\bibinfo
  {publisher} {Cambridge University Press},\ \bibinfo {address} {Cambridge},\
  \bibinfo {year} {2004})\BibitemShut {NoStop}%
\bibitem [{\citenamefont {Robnik}(1981)}]{Rob1981}%
  \BibitemOpen
  \bibfield  {author} {\bibinfo {author} {\bibfnamefont {M.}~\bibnamefont
  {Robnik}},\ }\href@noop {} {\bibfield  {journal} {\bibinfo  {journal} {J.
  Phys. A: Math. Gen.}\ }\textbf {\bibinfo {volume} {14}},\ \bibinfo {pages}
  {3195} (\bibinfo {year} {1981})}\BibitemShut {NoStop}%
\bibitem [{\citenamefont {Robnik}(1982)}]{Rob1982}%
  \BibitemOpen
  \bibfield  {author} {\bibinfo {author} {\bibfnamefont {M.}~\bibnamefont
  {Robnik}},\ }\href@noop {} {\bibfield  {journal} {\bibinfo  {journal} {J.
  Phys. Colloque C2}\ }\textbf {\bibinfo {volume} {43}},\ \bibinfo {pages} {29}
  (\bibinfo {year} {1982})}\BibitemShut {NoStop}%
\bibitem [{\citenamefont {Hasegawa}\ \emph {et~al.}(1989)\citenamefont
  {Hasegawa}, \citenamefont {Robnik},\ and\ \citenamefont {Wunner}}]{HRW1989}%
  \BibitemOpen
  \bibfield  {author} {\bibinfo {author} {\bibfnamefont {H.}~\bibnamefont
  {Hasegawa}}, \bibinfo {author} {\bibfnamefont {M.}~\bibnamefont {Robnik}}, \
  and\ \bibinfo {author} {\bibfnamefont {G.}~\bibnamefont {Wunner}},\
  }\href@noop {} {\bibfield  {journal} {\bibinfo  {journal} {Prog. Theor. Phys.
  Suppl. (Kyoto)}\ }\textbf {\bibinfo {volume} {98}},\ \bibinfo {pages} {198}
  (\bibinfo {year} {1989})}\BibitemShut {NoStop}%
\bibitem [{\citenamefont {Wintgen}\ and\ \citenamefont
  {Friedrich}(1989)}]{WF1989}%
  \BibitemOpen
  \bibfield  {author} {\bibinfo {author} {\bibfnamefont {D.}~\bibnamefont
  {Wintgen}}\ and\ \bibinfo {author} {\bibfnamefont {H.}~\bibnamefont
  {Friedrich}},\ }\href@noop {} {\bibfield  {journal} {\bibinfo  {journal}
  {Phys. Rep.}\ }\textbf {\bibinfo {volume} {183}},\ \bibinfo {pages} {38}
  (\bibinfo {year} {1989})}\BibitemShut {NoStop}%
\bibitem [{\citenamefont {Ruder}\ \emph {et~al.}(1994)\citenamefont {Ruder},
  \citenamefont {Wunner}, \citenamefont {Herold},\ and\ \citenamefont
  {Geyer}}]{RWHG1994}%
  \BibitemOpen
  \bibfield  {author} {\bibinfo {author} {\bibfnamefont {H.}~\bibnamefont
  {Ruder}}, \bibinfo {author} {\bibfnamefont {G.}~\bibnamefont {Wunner}},
  \bibinfo {author} {\bibfnamefont {H.}~\bibnamefont {Herold}}, \ and\ \bibinfo
  {author} {\bibfnamefont {F.}~\bibnamefont {Geyer}},\ }\href@noop {} {\emph
  {\bibinfo {title} {Atoms in Strong Magnetic Fields}}}\ (\bibinfo  {publisher}
  {Heidelberg: Springer},\ \bibinfo {year} {1994})\BibitemShut {NoStop}%
\end{thebibliography}
\providecommand{\noopsort}[1]{}\providecommand{\singleletter}[1]{#1}%

\end{document}